\documentclass[12pt]{article}

\usepackage[dvips]{graphicx}

\graphicspath{{Plot/}}

\newcommand{\lsim}{\raisebox{-4pt}{$\,\stackrel{\textstyle
                                                         <}{\sim}\,$}}
\newcommand{\gsim}{\raisebox{-4pt}{$\,\stackrel{\textstyle
                                                         >}{\sim}\,$}}
\newcommand{\nn}{\nonumber}
\newcommand{\be}{\begin{equation}}
\newcommand{\ee}{\end{equation}}
\newcommand{\ba}{\begin{eqnarray}}
\newcommand{\ea}{\end{eqnarray}}
\newcommand{\req}[1]{(\ref{#1})}
\def\={\,=\,}
\newcommand{\ci}[1]{\cite{#1}}

\def\gev{~{\rm GeV}}

\def\ale{\alpha_{\rm em}}
\def\als{\alpha_{\rm s}}

\newcommand{\tw}{\textwidth}

\def\vd{{\bf \Delta}_\perp}
\def\vbs{{\bf b}}
\def\vb0{{\bf b}_0}

\newcommand{\wf}{wave function}

\def\sh{\hat{s}}
\def\uh{\hat{u}}
\def\th{\hat{t}}

\begin{document} 
\thispagestyle{empty}
\begin{flushright}
WUB/17-01\\
March, 15  2017\\[20mm]
\end{flushright}

\begin{center}
{\Large\bf The GPD $\widetilde H$ and spin correlations in wide-angle
Compton scattering}
\vskip 10mm

P.\ Kroll \footnote{Email:  pkroll@uni-wuppertal.de}
\\[1em]
{\small {\it Fachbereich Physik, Universit\"at Wuppertal, D-42097 Wuppertal,
Germany}}\\

\end{center}
\vskip 5mm 
\begin{abstract}
Wide-angle Compton scattering (WACS) is discussed within the handbag approach in which 
the amplitudes are given by products of hard subprocess amplitudes and form factors, specific
to Compton scattering, which represent $1/x$-moments of generalized parton distributions (GPDs).
The quality of our present knowledge of these form factors and of the underlying GPDs
is examined. As will be discussed in some detail the form factor $R_A$ and the underlying GPD 
$\widetilde H$ are poorly known. It is argued that future data on the spin correlations $A_{LL}$
and/or $K_{LL}$ will allow for an extraction of $R_A$ which can be used to constrain the large
$-t$ behavior of $\widetilde{H}$.  
\end{abstract} 
\section{Introduction}
\label{sec:intro}
\begin{figure}[t]
\centering
\includegraphics[width=0.5\tw]{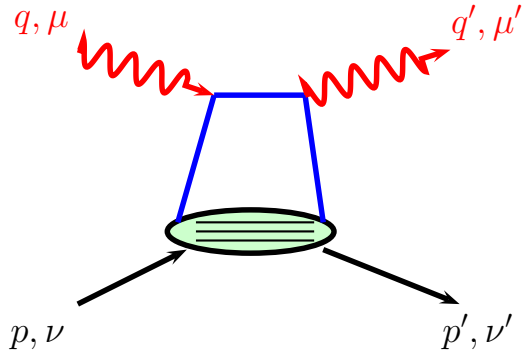}
\caption{\label{fig:handbag} The handbag contribution to WACS. The particle momenta and 
helicities are quoted. The horizontal lines represent any number of spectator partons. }
\end{figure}

Hard exclusive processes have extensively be investigated, experimentally as well
as theoretically, over the last twenty years. It is fair to say that some understanding
of these processes have been achieved so far. Thus, it is clear now that the handbag 
graph shown in Fig.\ \ref{fig:handbag}, controls the two complementary processes -
deeply virtual (DVCS) \ci{rad97,ji98,collins98} and wide-angle Compton scattering 
\ci{rad98, DFJK1} even for kinematics accessible at the Jefferson Lab. DVCS is
characterized by small momentum transfer from the initial to the final state proton 
and a large photon virtuality. The amplitudes in this case are given by convolutions
of hard subprocess amplitudes and soft proton matrix elements parametrized as GPDs. 
As derived in \ci{rad98, DFJK1}, for large Mandelstam variables, $s, -t, -u$, the WACS 
amplitudes are given by products of hard subprocess amplitudes and form factors, specific 
to WACS, which represent $1/x$-moments of GPDs. The handbag approach can be generalized to 
deeply virtual meson electroproduction (DVMP) and to wide-angle photoproduction of mesons. 
It turned out that in both cases the numerical estimations of cross sections fail by order 
of magnitude in comparison with experiment at least for Jefferson Lab kinematics 
\ci{GK2,mueller11,HK00}. For pion electroproduction an explanation of this failure has been 
found: lacking contributions from transversity GPDs, formally of twist-3 nature, which are 
strongly enhanced by the chiral condensate \ci{GK5,GK6,liuti}. Inclusion of these 
contributions leads to reasonable agreement with experiment \ci{clas-pi0,hallA}. Whether 
this mechanism also explains the underestimate of the pion-photoproduction cross section 
is not yet clear \ci{passek}. 
 
With regard to planned experiments at Jefferson Lab it seems timely to have a fresh look
at WACS within the handbag approach. As compared to the situation around 2000 there is
new aspect: we have now a fair knowledge of the GPDs $H$ and $E$ at large $-t$,
underlying the Compton form factors $R_V$ and $R_A$, respectively, from an analysis of the 
nucleon form factors \ci{DK13}. On the other hand, our present knowledge of the GPD 
$\widetilde H$ at large $-t$, related to the form factor $R_A$, is still poor due to the very 
limited experimental information available on the isovector axial form factor of the nucleon
in that kinematical range. It is important to realize that for known Compton form
factors, evaluated for instance from given GPDs, the WACS cross section as well as 
spin-dependent observables can be computed free of any adjustable parameter. However,
because of the poor knowledge of $\widetilde H$, the present numerical computations of the 
form factor $R_A$ suffer from large uncertainties and, therefore, predictions for WACS
observables which are sensitive to $R_A$, too. With regard to the interest in the GPD
$\widetilde H$ at large $-t$ for studying the impact parameter distribution of
quarks with definite helicity it will be proposed in this article to turn the strategy
around and to extract the form factor $R_A$ from spin correlations like $A_{LL}$ or
$K_{LL}$. Accurate data on such spin correlations at sufficiently large $-t$ and $-u$
will provide a set of values on $R_A(t)$ which subsequently can be used as a constraint
on $\widetilde H$ in addition to the data on the axial form factor. Also for this
form factor more and better data are to be expected  from the FNAL MINERVA experiment
in the near future. Constraining $\widetilde H$ by data on $F_A$ and $R_A$ will also allow 
for a more accurate flavor separation as was possible up to now. It even might be possible 
to say something about $\widetilde H$ for sea quarks at large $-t$. 

The plan of the paper is the following: In Sect.\ \ref{sec:handbag} the handbag approach
to WACS will be recapitulated and in Sect.\ \ref{sec:Htilde} the properties of the
zero-skewness GPDs at large $-t$ will be discussed. Sect.\ \ref{sec:spin} is devoted
to a discussion of spin correlations with regard to their sensitivity to $R_A$. A
summary is given in Sect.\ \ref{sec:summary}. In the appendices different conventions
for the spin observables are presented and compared to each other. 
 
\section{The structure of the handbag mechanism} 
\label{sec:handbag}
For the description of the handbag contribution to WACS we follow Ref.\ \ci{DFJK1}. 
It is assumed that the Mandelstam variables $s, -t$ and $-u$ are much larger than 
$\Lambda^2$ where $\Lambda$ is a typical hadronic scale of order $1\,\gev$. It is of 
advantage to work in a symmetric frame where the momenta of the initial ($p$) and final 
($p'$) state protons are parametrized as 
\be
  p^{(')} \= \Big[p^+\,,\frac{m^2-\vd^2/4}{2p^+}\,,{}^{\phantom{(}-}_{(+)} \frac12\vd\Big]
\ee    
in light-cone coordinates; $m$ denotes the proton mass. The photon momenta ($q$ and
$q'$) are defined analogously. In this frame the plus and minus light-cone components
of the momentum transfer, $\Delta=p' - p$, are zero implying $t=-\vd^2$ as well as
a vanishing skewness parameter $\xi=(p-p')^+/(p+p')^+$. The handbag contribution, shown in 
Fig.\ \ref{fig:handbag}, is then defined through the assumption that the soft hadronic
\wf s occurring in the Fock decomposition of the proton state are dominated by parton
virtualities in the range $|k_i^2|\lsim \Lambda^2$ and by intrinsic transverse
momenta, $k_{\perp i}$, that satisfy $k^2_{\perp i}/x_i\lsim \Lambda^2$ where 
$x_i=k_i^+/p^+$ is the usual light-cone momentum fraction.~\footnote{
   A restriction of transverse momenta by  $k^2_{\perp i}\lsim \Lambda^2$ instead fails
   to ensure small parton virtualities in the proton. At least one parton virtuality
   would be of order $\Lambda\sqrt{-t}$ and not $\Lambda^2$.}
In frames with non-zero skewness there are additional contributions. 
On these presuppositions one can show that the photon-parton scattering is hard and
the momenta of the active partons, $k_j, k'_j$, are approximately on-shell, collinear
with their parent hadrons and with momentum fractions, $x_j, x'_j$, close to 1. A
consequence of these properties is that the Mandelstam variables in the photon-parton
subprocess, $\sh, \th, \uh$, and in the overall photon-proton reaction, $s, t, u$, are
approximately equal up to corrections of order $\Lambda^2/t$. The propagator poles of
the handbag graphs at $\sh=0$ (in the graph shown in Fig.\ \ref{fig:handbag}) and $\uh=0$ 
(for the graph with the two photons crossed) are thus avoided. The physical picture is that
of a hard photon-parton scattering and soft emission and re-absorption of partons by the 
proton.

The (light-cone) helicity amplitudes for WACS in the symmetric frame are then given by 
\ci{DFJK1,HKM}
\ba
{\cal M}_{\mu'+,\mu +}(s,t)&=& 2\pi\ale \Big[{\cal H}_{\mu'+,\mu +}(\sh,\th)
       \big(R_V(t)+R_A(t)\big) \nn\\
            &&  + {\cal H}_{\mu'-,\mu -}(\sh,\th) \big(R_V(t)-R_A(t)\big)\,,\nn\\
{\cal M}_{\mu'-,\mu +}(s,t)&=& \pi\ale\frac{\sqrt{-t}}{m} \Big[{\cal H}_{\mu'+,\mu +}(\sh,\th)
                 + {\cal H}_{\mu'+,\mu +}(\sh,\th)\Big]\,R_T(t)\,.
\label{eq:amplitudes}
\ea
The amplitudes are subject to uncontrolled corrections of order $\Lambda^2/t$.
Explicit helicities are labeled only by their signs. In the Compton amplitudes, ${\cal M}$, 
the explicit helicities refer to those of the protons, in the subprocess amplitudes, 
${\cal H}$, to those of the active partons.

The soft proton matrix elements $R_i$ ($i=V, A, T$), appearing in \req{eq:amplitudes},
represent new types of proton form factors specific to Compton scattering. These Compton 
form factors are defined as $1/x$-moments of zero-skewness GPDs for 
$-t\gg \Lambda^2$. For active quarks of flavor $a$ ($u, d, \ldots$) they read~\footnote{
  Since we only consider zero-skewness GPDs their skewness argument is dropped unless stated
  otherwise.}~\footnote{
  For a discussion of the scale-dependence of the Compton form factors see \ci{DFJK4}.}
\ba
R_V^a(t) &=& \int_{-1}^1 \frac{dx}{x} H^a(x,t)\,, \nn\\
R_A^a(t) &=& \int_{-1}^1 \frac{dx}{x} {\rm sign}(x)\widetilde{H}^a(x,t)\,, \nn\\
R_T^a(t) &=& \int_{-1}^1 \frac{dx}{x} E^a(x,t)\,.
\label{eq:form-factors}
\ea
The full form factors in \req{eq:amplitudes} are given by the sum
\be
R_i(t) \= \sum_a e_a^2 R_i^a(t)\,,
\label{eq:compton-FF}
\ee
$e_a$ being the charge of the quark $a$ in units of the positron charge. In principle
there is a fourth form factor, related to the GPD ${\tilde E}$, which however decouples
in the symmetric frame. The flavor form factors \req{eq:form-factors} also appear
in wide-angle photoproduction of mesons \ci{HK00}.

The hard scattering amplitudes ${\cal H}_{\mu'\lambda',\mu\lambda}$ are to be calculated 
perturbatively. To leading order (LO), obtained from the graph shown in Fig.\ 
\ref{fig:handbag} and the one with the two photons crossed, they read
\be
{\cal H}^{\rm LO}_{++,++} \=2 \sqrt{\frac{\sh}{-\uh}}\,, \quad
{\cal H}^{\rm LO}_{++,++} \=2 \sqrt{\frac{-\uh}{\sh}}\,, \quad
{\cal H}^{\rm LO}_{-+,++} \=0\,.
\ee
Since the quarks are taken as massless there is no quark helicity flip to any order
of $\als$. In \ci{HKM} the next-to-leading order (NLO) corrections have also been calculated.
They provide phases and logarithmic corrections~\footnote{
    Since in WACS $-\th$ and $-\uh$ are of order $\sh$ there are no large logarithms
    in the NLO amplitudes.}
to the LO amplitudes and generate a non-zero photon helicity-flip amplitude
\be
{\cal H}^{\rm NLO}_{-+,++} \= -\frac{\als}{2\pi} C_F \left(\sqrt{\frac{\sh}{-\uh}}
                              + \sqrt{\frac{-\uh}{\sh}}\right)\,.
\label{eq:NLO-flip}
\ee

The matching of the subprocess and the full Mandelstam variables is simple if the mass of 
the proton can be neglected. In this case 
\be
{\mathrm scenario\; 1}:\hspace*{0.16\tw}  \sh\=s\,, \qquad \th\=t\,, \qquad \uh\=u\,.
\hspace*{0.20\tw}
\label{eq:scenario1}
\ee
In order to estimate the influence of the proton mass two more scenarios have been introduced
in \ci{DFHK03}: 
\ba
{\mathrm scenario\;2}: &&\hspace*{0.03\tw} \sh\=s-m^2\,, \qquad \th\=t\,, \qquad \uh\=u-m^2\,,\nn\\
{\mathrm scenario\;3}: &&\hspace*{0.03\tw} \sh\=s-m^2\,, \quad 
\th\=-\frac{\sh}{2}(1-\cos{\theta_{\rm cm}})\,, \quad \uh\=-\sh -\th\,.\quad
\label{eq:scenario3}
\ea
The center-of-mass-system (c.m.s.) scattering angle is denoted by $\theta_{\rm cm}$. In the latter 
two scenarios one has $\sh +\th +\uh=0$ in contrast to scenario 1. As advocated for in 
\ci{DFHK03} WACS observables are calculated in the three scenarios and the differences are 
considered as uncertainties of the predictions. 

To NLO there are also contributions from the gluonic subprocess $\gamma g\to\gamma g$.
They are in general small and we only take into account the most important one arising 
from the the gluonic GPD $H^g$:
\be
2\pi\ale \Big[{\cal H}^g_{\mu'+,\mu +} + {\cal H}^g_{\mu'-,\mu -}\Big]\,R_V^g(t)
\ee
which is to be added to the proton helicity non-flip amplitude in \req{eq:amplitudes}.
The explicit helicity labels refer to the gluon helicity now. The form factor $R^g_V$
is given by
\be
R_V^g(t)\=\sum_a e_a^2 \int_0^1\frac{dx}{x^2} H^g(x,t)\,.
\ee
The additional factor $1/x$ as compared to the form factors \req{eq:form-factors}
is conventional, it appears as a consequence of the definition of the gluon GPD whose
forward limit is
\be
xg(x) \= H^g(x,0)\,.
\ee
We refrain from quoting the NLO subprocess amplitudes here; they can be found in \ci{HKM}.

The last issue to be discussed in this section is the choice of the helicity basis. The
derivation of the amplitudes \req{eq:amplitudes} naturally requires the use of light-cone
helicities. However, for comparison with experiment the use of the ordinary helicities
is more convenient. Diehl \ci{diehl01} has given the transformation from on basis to the other. 
The standard helicity amplitudes, $\Phi_{\mu'\nu',\mu\nu}$, (the notation of the helicity labels 
are kept) are related to the light-cone amplitudes \req{eq:amplitudes} by
\ba
\Phi_{\mu'\nu',\mu\nu} &=& {\cal M}_{\mu'\nu',\mu\nu} + \frac{\eta}{2} \Big[
                       (-1)^{1/2-\nu'} {\cal M}_{\mu'-\nu',\mu\nu}    \nn\\
               && +  (-1)^{1/2+\nu} {\cal M}_{\mu'\nu',\mu-\nu} \Big] + {\cal O}(\Lambda^2/t)
\label{eq:transform}
\ea   
where
\be
\eta\=\frac{2m}{\sqrt{s}}\,\frac{\sqrt{-t}}{\sqrt{s} + \sqrt{-u}}\,.
\label{eq:transform-parameter}
\ee
In principle there are 16 amplitudes for Compton scattering. However parity and time-reversal
invariance lead to relations among them~\footnote{
Analogous relations for the other set of amplitudes, ${\cal M}$ and ${\cal H}$.}
\be
\Phi_{-\mu'-\nu',-\mu-\nu}\=\Phi_{\mu\nu,\mu'\nu'}\=
                           (-1)^{\mu-\nu-\mu'+\nu'}\,\Phi_{\mu'\nu',\mu\nu}\,.
\label{eq:parity}
\ee
With the help of \req{eq:parity} one sees that there are only 6 independent amplitudes 
for which we choose \ci{rollnik}
\ba
\Phi_1&=&\Phi_{++,++}\,, \quad \Phi_3\=\Phi_{-+,++}\,, \quad \Phi_5\=\Phi_{-+,-+}\,, \nn\\
\Phi_2&=&\Phi_{--,++}\,, \quad \Phi_4\=\Phi_{+-,++}\,, \quad \Phi_6\=\Phi_{-+,+-}\,.
\label{eq:cms-amplitudes}
\ea
Inspection of \req{eq:amplitudes} and \req{eq:transform} reveals that
\be
\Phi_2\=-\Phi_6 + {\cal O}(\Lambda^2/t)\,.
\label{eq:phi2-phi6}
\ee
This relation is a robust property of the handbag mechanism which is difficult to 
change. The photon helicity-flip amplitudes $\Phi_i$, $i=2,3,6$ are related to 
${\cal H}_{-+,++}^{\rm NLO}$ and, hence, they are of order $\als$ (see \req{eq:NLO-flip}).
\section{The GPDs at large $-t$}
\label{sec:Htilde}
The sum rules for the Dirac ($i=1$) and Pauli ($i=2$) form factors of the nucleon read
\be
F_i^{p(n)}(t)\=e_u F_i^{u(d)}(t) + e_d F_i^{d(u)}(t)
\label{eq:sum-rule}
\ee
with the flavor form factors defined by
\be
F_i^a(t)\=\int_0^1 dx K_{iv}^a(x,t)
\ee
where $K^a_{iv}$ denotes the relevant proton GPD, either $H^a_v$ for the Dirac form
factor or $E_v^a$ for the Pauli one. A valence-quark GPD is defined by the combination
\be
K^a_{iv}(x,t)\=K_i^a(x,t) + K_i^a(-x,t)\,.
\ee 
In \ci{DK13} the GPDs $H$ and $E$ for valence quarks have been extracted from the
data on the magnetic form factor of the proton and the neutron and from the ratios
of electric and magnetic form factors exploiting the sum rules \req{eq:sum-rule} with 
the help of a parametrization of the zero-skewness GPDs:
\be
K_{iv}^a(x,t)\=k_i^a(x) \exp{[tf_i^a(x)]}\,.
\label{eq:gpd}
\ee
In \ci{DK13,DFJK4} it is advocated for the following parametrization of the profile function
\be
f_i^a(x) \= \big(\alpha'_i{}^a\ln{(1/x)} + B_i^a\big)(1-x)^3 + A^a_i x(1-x)^2
\label{eq:profile}
\ee
which differs from the Regge-like parametrization 
\be
f_{iR}^a(x) \= \alpha'_i{}^a\ln{(1/x)} + B_i^a
\label{eq:regge-profile}
\ee
frequently used in DVCS and DVMP. The above parametrization refers to a definite factorization 
scale $\mu_F$ for which we take $2\,\gev$ throughout this work. The forward limit of the GPD $H^a$ 
is given by the flavor-$a$ parton densities, $q^a(x)$, for which the ABM11 densities  \ci{ABM11}, 
evaluated at the scale $\mu_F$, are used in \ci{DK13}. The forward limit of $E^a$ which is not 
accessible in deep-inelastic scattering and is, therefore, to be determined in the form factor 
analysis, too. It is parametrized like the parton densities
\be
e^a(x)\=N_a x^{\alpha_e^a}(1 - x)^{\beta^a_e}(1 + \gamma^a\sqrt{x})\,.
\label{eq:Eforward}
\ee
The normalization, $N_a$, is obtained from the contribution of quarks of flavor $a$ to
the anomalous magnetic moment of the nucleon ($\kappa_u=1.67$, $\kappa_d=-2.03$)
\be
\kappa_a\=\int_0^1 dx E^a(x,0)\,.
\ee
The parameters of the profile functions as well as the additional ones for $e^a(x)$ are
fitted to the nucleon form factor data in \ci{DK13} and from the resulting GPDs the
Compton form factors $R_V$ and $R_T$ are subsequently evaluated. These form factors will be 
used in the following without exception.

The sum rule for the isovector axial form factor reads
\ba
F_A(t)&=&\int_0^1 dx \big[{\widetilde H}^u_v(x,t) - {\widetilde H}^d_v(x,t)\big] \nn\\
        &+&  2\int_0^1 dx\big[{\widetilde H}^u(-x,t) - {\widetilde H}^d(-x,t)\big]\,.
\ea
In contrast to the electromagnetic form factors the sea quark contributions do not drop in 
this case. The GPD $\widetilde H$ for valence quark is parametrized in the same fashion as 
$H$ with the unpolarized parton densities replaced by the polarized ones
\be
{\widetilde H}^u_v(x,t) \= \Delta q^a(x) \exp{[t{\tilde f}^a(x)]}\,.
\label{eq:htilde}
\ee 
The profile function is parametrized as in \req{eq:profile}. In \ci{DK13} a fit of the parameters 
for $\widetilde H$ has not been attempted since the data basis for the axial form factor is meager. 
The only set of data that covers a fairly large range of $-t$ ($\leq 3\,\gev^2$) is that one 
measured by Kitagaki {\it et al} \ci{kita83}. These data are presented in a form of a dipole fit 
with a mass parameter $M_A=(1.05^{+0.12}_{-0.16})\,\gev$ and are shown in Fig.\ \ref{fig:FA}. More 
accurate data on $F_A$ are to be expected from the FNAL MINERVA experiment. Therefore, with regard 
to the current data situation, it has been assumed in \ci{DK13} that $\tilde{f}^a$ is the same as 
the profile function, $f_1^a$, for $H^a_v$. The sea quark contribution has been neglected. It is very 
small at large $-t$ as estimated in \ci{DK13} (see also the discussion below). The polarized parton 
densities of \ci{DSSV09} were taken for the forward limit at the scale $2\,\gev$. The results on 
$F_A$ and $R_A$ are displayed in Figs.\ \ref{fig:FA} and \ref{fig:RA}, respectively. The axial form 
factor, although in agreement with experiment, lies at the lower edge of the data. The parameters of 
the profile function ${\tilde f}^a$ used in \ci{DK13} are quoted in Tab.\ \ref{tab:1}.
\begin{figure}[t]
\centering
\includegraphics[width=0.47\tw]{fig-FA-new.epsi}
\includegraphics[width=0.47\tw]{fig-RA-new.epsi}
\caption{\label{fig:FA} Left: The axial form factor, scaled by $-t$, versus $\sqrt{-t}$.  The 
    data \ci{kita83} are shown as the broad shaded band. The narrow shaded band represents the 
    results of \ci{DK13}. The solid (dashed) line is the example \#1 (\#2).}
\caption{\label{fig:RA} Right: The Compton form factor $R_A$, scaled by $t^2$, versus $\sqrt{-t}$.
    For further notations it is refered to Fig.\ \ref{fig:FA}.}
\end{figure} 

Let us consider the Fourier transformation of the zero-skewness valence quark GPDs to the 
impact parameter plane  
\be
k_i^a(x,b^2)\=\int \frac{d^2\vd}{(2\pi)^2} e^{-i\vbs\cdot\vd} K_{iv}^a(x,t=-\vd^2)\,.
\ee
Explicitly the parametrization \req{eq:gpd} leads to
\be
k_i^a(x,b^2)\=\frac1{4\pi}\frac{k_i^a(x)}{f_i^a(x)}\,\exp{[-\frac{b^2}{4f_i^a(x)}]}\,.
\ee
The sum and difference of the unpolarized and the longitudinally polarized 
impact parameter distributions 
\be
q_v^a(x,b^2) \pm \Delta q_v^a(x,b^2)
\ee
possess a density interpretation \ci{burkardt02} which implies the bound \ci{DFJK4}
\be
{\tilde f}^a(x)\leq f^a(x)
\label{eq:bound}
\ee
in the region where antiquarks can be neglected. Taking ${\tilde f}^a(x)<f^a(x)$ instead of
${\tilde f}^a(x)=f^a(x)$ increases the flavor form factor 
\be
{\tilde F}^a(t)\=\int_0^1 dx {\widetilde H}_v^a(x,t)\,,
\label{eq:flavorFA}
\ee
in particular at large $-t$. As we will discuss in Sect.\ \ref{sec:spin} the data on the 
helicity correlation $K_{LL}$ \ci{hallA05,hallA15}, although measured at values of $-t$ or 
$-u$ being not sufficiently large for an application of the handbag approach, have rather 
large values. This may be taken as a hint at larger values of $R_A$ than quoted in \ci{DK13}, 
see Fig.\ \ref{fig:RA}.

In order to understand the reason for choosing the complicated profile function \req{eq:profile} 
let us discuss the general behavior of the GPDs \req{eq:gpd}. They behave Regge-like for 
$x\to 0$, i.e.\ for ${\widetilde H}$~\footnote{
    An analogous discussion can be established for the other GPDs.}
\be
{\widetilde H}^a \sim x^{-\tilde{\alpha}_a -t \tilde{\alpha}'_a}
\ee
where the power $\tilde{\alpha}_a$, the intercept of a Regge-like trajectory, is hidden in the 
polarized parton densities. For $u$-valence quarks its value is about 0.1 as obtained from a fit
to the DSSV PDFs \ci{DSSV09} in the range $10^{-3} <x< 10^{-2}$. $\Delta q^d_v(x)$ behaves similar 
but with very large uncertainties. I.e.\, at small $-t$, the GPDs are singular for $x\to 0$ as 
the parton densities. With increasing $-t$ the singularity becomes milder and turns into a zero for 
\be
-t > -t_0 \= \tilde{\alpha}_a/\tilde{\alpha}'_a\,.
\ee
With $\tilde{\alpha}'_a\simeq 0.9\,\gev^{-2}$ $t_0$ is about $-0.1\,\gev^2$. While the flavor 
form factors \req{eq:flavorFA} exist for all $t$, the $1/x$-moments are singular for $-t<-t_0$. 
This is unproblematic since the Compton form factors are only defined for$-t\gg \Lambda^2$. In 
order to achieve larger values for $R_A$ at intermediate $-t$ it is plausible to use smaller 
values of $\tilde{\alpha}'_a$ and, hence, a larger value of $-t_0$ closer to the $t$-range of 
applicability of the handbag approach. For this purpose we consider two examples: For the case 
\#1 we take $\simeq\tilde{\alpha}'_a/2$, characteristic of a Regge cut, and for case \#2 
$\tilde{\alpha}'_u=0.144 $ but leaving $\tilde{\alpha}'_d$ unchanged, see Tab.\ \ref{tab:1}. In 
both cases the other parameters in the profile function are fitted to the $F_A$ data ( for case \#1
common factors for the $u$ and $d$ quark parameters $\tilde A$ and $\tilde B$ are used). The 
parameters of $\tilde{f}^a$ are compiled in Tab.\ \ref{tab:1}. In all cases the bound 
\req{eq:bound} is respected for all $x$.    
\begin{table*}[h]
\renewcommand{\arraystretch}{1.4} 
\begin{center}
\begin{tabular}{| c || c  c  c | c  c  c|}
\hline   
  &  &  ${\tilde f}^u$ & & & ${\tilde f}^d$ &  \\[0.2em]
  & $\tilde{\alpha}'_u$ & $\tilde{B}_u$ & $\tilde{A}_u$ & $\tilde{\alpha}'_d$ & $\tilde{B}_d$ 
& $\tilde{A}_d$ \\[0.2em]
\hline
 \ci{DK13} & 0.961 & 0.545 & 1.264 & 0.861 & 0.333 & 4.198 \\[0.2em]
     \#1   & 0.432 & 0.654 & 1.239 & 0.387 & 0.400 & 4.284 \\[0.2em]
     \#2   & 0.144 & 1.100 & 1.150 & 0.861 & 0.333 & 4.198 \\[0.2em]
\hline
\end{tabular}
\end{center}
\caption{Parameters of the profile function ${\tilde f}$ for $u$ and $d$ 
valence quarks. All quantities are given in units of $\gev^{-2}$.}
\label{tab:1}
\renewcommand{\arraystretch}{1.0}   
\end{table*} 

The results on $F_A$ for the cases \#1 and \#2 are also shown in Fig.\ \ref{fig:FA}. Both 
the examples are well in agreement with the data but substantially larger for $\sqrt{-t}$
in the range $1 - 2\,\gev$ than the form factor proposed in \ci{DK13} from the assumption 
$\tilde{f}^a=f_1^a$for $\sqrt{-t}\simeq 2\,\gev$. At large $\sqrt{-t}$ all 
three cases are very close to each other. A similar behavior is seen in Fig.\ \ref{fig:RA} 
for the form factor $R_A$.   The GPD ${\widetilde H}^a$ for the three cases are displayed 
in Fig.\ \ref{fig:Htilde}. The GPD exhibits a pronounced maximum (minimum) at a large value
of $x$, its position moves towards larger values of $x$ for increasing $-t$ and becomes 
narrower. For the three cases $\widetilde H$ looks very similar, there are only small 
differences in the position and the height of the maximum respective minimum. The above 
considerations tell us that there is a strong $x - t$ correlation in the GPDs \req{eq:gpd}.
\begin{figure}[t]
\centering
\includegraphics[width=0.45\tw]{fig-Htilde-new.epsi}
\includegraphics[width=0.48\tw]{fig-b-distribution-ratio.epsi}
\caption{\label{fig:Htilde} Left: The zero-skewness GPD $\widetilde H$ versus $x$ at the scale
         $2\,\gev$. The solid (dashed, dotted) line represents the result obtained from \ci{DK13} 
         (\#1, \#2).} 
\caption{\label{fig:b-distribution} Right: The ratio of $q_v^u(x,b^2)$ for the Regge-like profile
function \req{eq:regge-profile} and for \req{eq:profile} at $x=0.05 (0.2, 0.6)$ solid 
(dashed, dotted) line. The scale is $2\,\gev$.}
\end{figure} 

Due to the $x-t$ correlation the flavor form factors calculated from the GPDs  
\req{eq:gpd}, \req{eq:htilde} are under control of large $x$ at large $-t$.
The polarized and unpolarized parton densities behave as $(1-x)^\beta$ for $x\to 1$ and an 
analogous behavior holds for $E$, see \req{eq:Eforward}. For valence quarks the powers 
$\beta$ are obtained from fits to the large $x$ behavior ($0.65 < x < 0.85$) of the ABM 
\ci{ABM11} and DSSV \ci{DSSV09} PDFs. The corresponding powers of $E$ are determined in 
the nucleon form factor analysis performed in \ci{DK13}. The various powers of the 
valence-quark GPDs are compiled in Tab.\ \ref{tab:2}. With the help of the saddle point 
method one can show \ci{DK13} that the flavor form factors behave as~\footnote{
   A power law behavior of form factors and other exclusive observables have also been obtained
   from soft physics, namely from overlaps of light-cone \wf s, in \ci{ralston}.} 
\be
 F_i(t) \sim  1/(-t)^{(1+\beta)/2}
\label{eq:power-law}
\ee
at sufficiently large $-t$ (the powers $(1+\beta/2)$ are also quoted in Tab.\ \ref{tab:2}). 
One also sees that the saddle point lies in the so-called soft region
\be
1-x \sim \Lambda/\sqrt{-t}
\ee
where, again at sufficiently large $-t$, the active parton carries  most of the proton's
momentum while all spectators are soft. This is the region of the Feynman mechanism, 
discussed already by Drell and Yan \ci{DY}. 

The GPD parametrizations \req{eq:gpd} and \req{eq:htilde} may analogously be extended to
sea quarks and gluons. For the corresponding flavor form factors, i.e.\ their lowest moments, 
the power behavior \req{eq:power-law} holds too. The powers $\beta$ of the PDFs for sea 
quarks and gluons are $\gsim 5$. This is in agreement with perturbative QCD considerations 
in the limit $x\to 1$ \ci{BBS95}. Similar powers are expected for $E$ for gluons and sea 
quarks. The corresponding flavor form factors are therefore strongly suppressed, the 
valence-quark form factors dominate at large $-t$. Since for $x\to 1$ the $1/x$-factor in 
the Compton form factors can be neglected, these form factors are also dominated by the 
valence quark contributions and in particular by the $u$-valence quark one. This has the 
following consequences which hold at sufficiently large 
$-t$:
\be
R_V \simeq e_u^2 F_1^u\,, \qquad R_A\simeq e_u^2 {\tilde F}^u\,, \qquad R_T/R_V\simeq 1/|t|^{0.6}\,,
\ee 
and for the ratio of the form factors for Compton scattering off neutrons and off protons
one has
\be
R_i^n/R_i^p \simeq e_d^2/e_u^2\,.
\ee
\begin{table*}[h]
\renewcommand{\arraystretch}{1.4} 
\begin{center}
\begin{tabular}{| c || c | c| c | c | c | c |}
\hline   
   &  $H^u_v$  & $H^d_v$ &  $E^u_v$ & $E^d_v$ &${\widetilde H}^u_v$ & ${\widetilde H}^d_v$  \\[0.2em]
\hline
$\beta$       & 3.50  & 5.00 & 4.65 & 5.25 & 3.43 & 4.22 \\[0.2em]
$(1+\beta)/2$ & 2.25  & 3.00 & 2.83 & 3.12 & 2.22 & 2.61  \\[0.2em]
\hline
\end{tabular}
\end{center}
\caption{The powers $\beta$ and $(1+\beta)/2$ for the valence quark GPDs.}
\label{tab:2}
\renewcommand{\arraystretch}{1.0}   
\end{table*} 

Another issue is the role of the third term in \req{eq:profile} which in fact is responsible
for the large $-t$ behavior of the GPDs. This term cannot be fixed from DVCS and DVMP 
data since they are typically measured at rather small values of $-t$ (note that 
factorization of these processes requires $-t \ll Q^2$). Therefore, the Regge-like profile 
function \req{eq:regge-profile} is frequently used in the analysis of these processes. As 
discussed in \ci{DFJK4} the Regge-like profile function although it is a reasonable 
approximation at low $x$ (low $-t$), is unphysical at large $x$ (large $-t$).
The transverse distance between the active parton and the cluster of spectators,
say for the GPD $H$, is given by
\be
d_a(x) \= \frac{\sqrt{\langle b^2 \rangle_x^a}}{1-x}\= 2\frac{\sqrt{f_1^a(x)}}{1-x}
\ee
for the parametrization \req{eq:gpd}. In the limit $x\to 1$, the profile function
\req{eq:profile} leads to a finite result for the distance $d_a(x)$ ($\sim 2 \sqrt{A^a}$)
while for \req{eq:regge-profile} the distance $d_a$ becomes singular ($\sim 2\sqrt{B^a}/(1-x)$).
This consideration makes it clear that the Regge-like profile function does not allow
for an investigation of the localization of the partons in the impact parameter plane.
This is also obvious from Fig.\ \ref{fig:b-distribution} where the ratio of $q^u_v(x,b^2)$ 
evaluated from the profile functions \req{eq:regge-profile} and \req{eq:profile}
is displayed for several values of $x$. While there is not much difference between
the two distributions, $q^u_{\rm R}(x, b^2)$ (evaluated from \req{eq:regge-profile}) and 
$q^u_{\rm DK}(x, b^2)$ (evaluated from \req{eq:profile}) at small $x$,
the Regge-like profile function leads to a much wider distribution at large $x$ than
the profile function \req{eq:profile}. 

\begin{figure}[t]
\centering
\includegraphics[width=0.45\tw]{fig-b-distribution-Htilde-0.05.epsi}
\includegraphics[width=0.46\tw]{fig-b-distribution-Htilde-0.6.epsi}
\caption{\label{fig:b-distribution-0.05} Left: The impact parameter distributions for valence
    quarks with definite helicities at $x=0.05$ (in fm${}^{-2}$). $\Delta q(x,b^2)$ is evaluated 
    from the profile function for example \#1. The PDFs are taken at the scale $2\,\gev$.} 
\caption{\label{fig:b-distribution-0.6} Right: as Fig.\ \ref{fig:b-distribution-0.05} but for
      $x=0.6$.}
\end{figure} 
Finally, we want to discuss the impact-parameter distribution of valence quarks with
definite helicities, defined by
\be
q^a_{v\pm}(x,b^2)\=\frac12\Big[ q^a_v(x,b^2) \pm \Delta q^a_v(x,b^2)\Big]\,.
\ee
Here, $q_{\pm}(x,b^2)$ is the distribution of quarks with helicity parallel ($+$) or 
anti-parallel ($-$) to the proton's helicity. These distributions are axially symmetrical 
around the direction of the proton momentum. In Figs.\ \ref{fig:b-distribution-0.05} and 
\ref{fig:b-distribution-0.6}, we show these distributions versus $b$ at $x=0.05$ and $0.6$,
respectively. By far the most important distribution is the $u$-quark one with parallel 
helicity. At large $x$ in particular the other three distributions can be neglected. At low 
$x$ the second largest distribution is the $d$-quark one with anti-parallel helicity. The 
dominance of the $u$-quark distribution with parallel helicity at large $x$ is expected from 
perturbative QCD \ci{BBS95}. On the other hand, the behavior of the $d$-quark distribution 
does not match the perturbative QCD predictions at the current experimentally accessible 
range of $x$ - the helicity-parallel distribution does not dominate since $\Delta q^d$ is 
negative. One also observes from Figs.\ \ref{fig:b-distribution-0.05} and 
\ref{fig:b-distribution-0.6}  the typical behavior of the impact-parameter distribution: a 
very broad distribution at low $x$ which becomes narrower with increasing $x$, i.e.\ for 
$x\to 1$ the active parton is close to the proton's center of momentum \ci{burkardt02}.

\section{Spin correlations}
\label{sec:spin}
All numerical results for WACS observables are evaluated to order $\als$ and terms 
of order $\eta^2$ are neglected throughout. The Compton form factors $R_V$ and $R_T$ are
taken from \ci{DK13}; for the third form factor, $R_A$, the three examples are considered
which are shown in Fig.\ \ref{fig:RA} and for which the different profile functions for 
$\widetilde H$ are quoted in Tab.\ \ref{tab:1}. The gluonic form factor $R_V^g$ is also 
taken into account in addition to the valence quark Compton form factors. Numerical values 
for this form factor are taken from \ci{HK00} where this form factor is modeled as a 
light-cone wave function overlap.

The unpolarized differential cross section 
\be
\frac{d\sigma}{dt}\=\frac1{32\pi(s-m^2)^2}\Big[|\Phi_1|^2 + |\Phi_2|^2 + 2 |\Phi_3|^2
                           + 2 |\Phi_4|^2 + |\Phi_5|^2 + |\Phi_6|^2\Big]
\ee
evaluated for the two examples of $R_A$, \#1 and \#2, does not differ much from the result 
given in \ci{DK13} since the $R_A$ contribution is suppressed by $t^2/(s-u)^2$ in comparison 
to the $R_V$ one. We therefore refrain from showing results on the cross section but refer 
to \ci{DK13} and concentrate ourselves in this article on spin effects.~\footnote{
    Tables with numerical results for the observables can be obtained from the author on 
    request.}  
For later use we also quote the LO handbag result for the cross section:
\be
\frac{d\sigma}{dt}\=\frac{\pi\ale^2}{(s-m^2)^2}\,\frac{(s-u)^2}{-us}\,(1+\kappa_T^2)R_V^2(t)\,
            \Big[1 + \frac{t^2}{(s-u)^2}\,\frac{R_A^2(t)}{R_V^2(t)}\frac{1}{1+\kappa_T^2}\Big]  
\ee
where 
\be
\kappa_T(t)\=\frac{\sqrt{-t}}{2m}\,\frac{R_T(t)}{R_V(t)}\,.
\ee
This quantity aquires values of between about 0.3 and 0.6 and depends on $t$ only mildly 
at large $-t$, see Tab.\ \ref{tab:2}.
 
The first observables we are going to discuss is the helicity ($L$-type, see App.\ 
\ref{sec:HKM}) correlation between the initial state photon and proton defined in terms 
of cross sections $d\sigma(\mu\nu)$ by
\be
A_{LL}\= \frac{d\sigma(++)-d\sigma(+-)}{d\sigma(++)+d\sigma(+-)}\,,
\ee
leading to
\be
A_{LL} \frac{d\sigma}{dt}\=\frac1{32\pi(s-m^2)^2}\Big[|\Phi_1|^2 + |\Phi_2|^2
                          - |\Phi_5|^2 - |\Phi_6|^2\Big]
\ee
in terms of the helicity amplitudes \req{eq:cms-amplitudes}. The analogous 
correlation between the helicities of the incoming photon and the outgoing proton
reads
\be
K_{LL}\=\frac{d\sigma(++)-d\sigma(+-)}{d\sigma(++)+d\sigma(+-)}
\ee
in terms of the cross sections $d\sigma(\mu\nu')$. Expressed through the c.m.s. helicity 
amplitudes it reads
\be
K_{LL} \frac{d\sigma}{dt}\=\frac1{32\pi(s-m^2)^2}\Big[|\Phi_1|^2 - |\Phi_2|^2
                          - |\Phi_5|^2 + |\Phi_6|^2\Big]\,.
\ee
Since $\Phi_2=-\Phi_6$, see \req{eq:phi2-phi6}, one arrives at
\be
A_{LL}\=K_{LL}\,.
\ee 
As mentioned in Sect.\ \ref{sec:handbag} this is a robust prediction of the handbag approach. 
However, using massive point-like quarks, $A_{LL}$ and $K_{LL}$ differ from each other  
in the backward hemisphere where $A_{LL}$ becomes smaller than $K_{LL}$ \ci{DFHK03,miller04}. 
Handbag results on $A_{LL}=K_{LL}$ are shown in Fig.\ \ref{fig:KLL} in a kinematical region 
where $-t$ and $-u$ are at least larger than about $2.5\,\gev^2$.~\footnote{
      It can be shown that in this kinematical range the present cross section data \ci{danagoulian07}
      are compatible with factorization of the RCS amplitudes in subprocess amplitudes and 
      form factor which only depend on $t$.}
The bands in the plot represent the results evaluated from example \#1 and are 
displayed for several values of $s$. The widths of the bands indicate possible kinematical 
corrections due to the mass of the proton \ci{DFHK03}. For $s$ below about $10\,\gev^2$ the 
uncertainty due to the proton mass corrections are rather large but become tolerable above 
$10\,\gev^2$. Results on  $A_{LL}=K_{LL}$ evaluated from example \#2 and from $R_A$ as quoted 
in \ci{DK13} are also shown at $s=7.8$ and $15\,\gev^2$. At fixed $t$ a strong energy 
dependence is to be noticed. The available data on $K_{LL}$ are also displayed in Fig.\ 
\ref{fig:KLL}. They are measured at c.m.s. scattering angles of $\theta_{\rm cm}=70^\circ$ 
\ci{hallA15} and $120^\circ$\ci{hallA05} at $s=7.8$ and $6.9\,\gev^2$, respectively. Both 
the data points are not compatible the prerequisite for the application of the handbag 
approach, namely $s,-t,-u\gg\Lambda^2$. For the data point at $70^\circ$ $t$ is 
$-2.1\,\gev^2$ and at $120^\circ$ $u$ is only $-1.04\,\gev^2$. 
\begin{figure}[t]
\centering
\includegraphics[width=0.45\tw]{fig-KLL-new-3.epsi}\hspace*{0.03\tw}
\includegraphics[width=0.47\tw]{fig-KLS-new-3.epsi}
\caption{\label{fig:KLL} Left: The helicity correlation $A_{LL}=K_{LL}$ versus $\sqrt{-t}$.
Uncertainty bands due to target mass corrections for fit 1 (upper edge: scenario 1; lower 
edge: scenario 3, see Eqs.\ \req{eq:scenario1}, \req{eq:scenario3}). Solid (dashed) lines 
for scenario 1, fit \#2 (\ci{DK13}) at $s=7.8$ (upper lines) and  $15\,\gev^2$ (lower lines). 
Data from \ci{hallA15} (solid circle) and \ci{hallA05} (open circle).} 
\caption{\label{fig:KLS} Right: The helicity correlation $A_{LS}=-K_{LS}$ versus $\sqrt{-t}$.
For notations it is refered to Fig.\ \ref{fig:KLL}.} 
\end{figure}

As one sees from the plot the helicity correlation parameter is very sensitive to the actual
value of $R_A$. It seems possible, as our analysis reveals, to achieve values for of $K_{LL}$ 
as large as the data \ci{hallA15,hallA05} indicate. The sensitivity of $A_{LL}$ and $K_{LL}$ 
on the form factor $R_A$, or, strictly speaking, on the ratio $R_A/R_V$ is obvious from a 
comparison with the LO result:
\be
K_{LL}\=2\frac{-t}{s-u}\,\frac{R_A}{R_V}\,\frac{1+\eta\kappa_T}{1+\kappa_T^2}
             \Big[1 + \frac{t^2}{(s-u)^2}\,\frac{R_A^2}{R_V^2}\frac1{1+\kappa_T^2}\Big]^{-1}\,.
\label{eq:ALL}
\ee
In a somewhat rough approximation the helicity correlation is given by the
Klein-Nishina helicity correlation for massless quarks 
\be
A_{LL}^{\rm KN}\= K_{LL}^{\rm KN}\=\frac{s^2-u^2}{s^2+u^2}
\label{eq:KN}
\ee 
diluted by the ratio of axial-vector over vector form factor, $R_A/R_V$. Hence, accurate
data on $A_{LL}$ and/or $K_{LL}$ would allow for a determination of that ratio and, 
subsequently, for an extraction of $R_A$ for a given vector form factor.

Somewhat similar is the correlation between the helicity of the 
incoming proton and the sideway polarization ($S$-type, see App.\ \ref{sec:HKM}) of the 
incoming proton defined by the following ratio of the cross sections 
$d\sigma(\mu\rightarrow)$ 
\be
A_{LS}\=\frac{d\sigma(+\rightarrow) - d\sigma(-\rightarrow)}
             {d\sigma(+\rightarrow) + d\sigma(-\rightarrow)}
\ee
which can be expressed as
\be
A_{LS} \frac{d\sigma}{dt}\=\frac1{16\pi(s-m^2)}\,{\rm Re}\,\Big[
            \big(\Phi^*_1 - \Phi^*_5\big)\Phi_4 - \big(\Phi^*_2 + \Phi^*_6\big)\Phi_3\Big]\,.
\ee
For the correlation between the helicity of the incoming photon and the sideway
polarization of the outgoing proton we analogously find
\be
K_{LS} \frac{d\sigma}{dt}\=\frac1{16\pi(s-m^2)}\,{\rm Re}\,\Big[
            \big(\Phi_1 - \Phi_5\big)\Phi_4^* + \big(\Phi_2 + \Phi_6\big)\Phi_3^*\Big]\,.
\ee
With $\Phi_2=-\Phi_6$ we have
\be
A_{LS}\=-K_{LS}
\ee
in the handbag approach.~\footnote{
      As explained in App.\ \ref{sec:BGLMN} a different convention is chosen in \ci{BGLMN}:
      A lab frame is used and the $L$ and $S$ spin directions for the incoming proton are
      opposite to the ones described in App.\ \ref{sec:HKM}.}
 The LO result for $K_{LS}$ reads:
\be
K_{LS}\=2\frac{-t}{s-u}\,\frac{R_A}{R_V}\,\frac{\kappa_T - \eta}{1+\kappa_T^2}
           \Big[1 + \frac{t^2}{(s-u)^2}\,\frac{R_A^2}{R_V^2}\frac1{1+\kappa_T^2}\Big]^{-1}\,.
\label{eq:KLS}
\ee
Thus, as $K_{LL}$, it is roughly given by the Klein-Nishina helicity correlation \req{eq:KN}
diluted by the ratio of $R_A$ and $R_V$ but additionally multiplied by the factor 
$\kappa_T - \eta$. This factor, for which $\kappa_T$ arises from the proton helicity 
flip amplitude \req{eq:amplitudes} and $\eta$ from the change of the helicity basis  
\req{eq:transform-parameter}, makes $K_{LS}$ very small for $s$ of about $8\,\gev^2$. 
For larger $s$ $K_{LS}$ becomes large in particular at large $-t$ as can be seen from Fig.\ 
\ref{fig:KLS}. In general the factor $\kappa_T-\eta$ makes $K_{LS}$ less suitable for an 
extraction of $R_A$ than $K_{LL}$ or $A_{LL}$. The predictions for $A_{LS}=-K_{LS}$ evaluated 
from the three examples quoted in Tab.\ \ref{tab:2} are shown in Fig.\ \ref{fig:KLS} and 
compared to the data \ci{hallA15}. The data point published in \ci{hallA05} is not shown in 
the plot. Its value is
\be
K_{LS}\=0.114 \pm 0.078 \pm 0.04  \qquad s\=6.9\,\gev^2 \qquad t\=-4.0\,\gev^2
\ee
and is a bit more than $1\sigma$ away from the prediction.

Many more spin correlation observables can be defined. Most of them are difficult to measure.
Several of them are zero due to parity invariance, e.g.\ $A_{LN}, A_{\parallel N}, A_{\perp N}$ 
(and the analogous $K_{ij}$ observables). Others are of order $\als$. An example of this is 
the correlation between a linearly polarized incoming photon, perpendicular to the scattering 
plane, and an $N$-type polarization of the incoming proton
\be
A_{\perp N}\= \frac{d\sigma(\perp\uparrow) - d\sigma(\perp\downarrow)}
                 {d\sigma(\perp\uparrow) + d\sigma(\perp\downarrow)} 
\ee
which in terms of helicity amplitude reads
\be
A_{\perp N}\frac{d\sigma}{dt} \= \frac1{16\pi(s-m^2)^2}\,{\rm Im}\Big[
                         (\Phi_1^* + \Phi_3^*)(\Phi_2 + \Phi_4)
                       + (\Phi_3^* + \Phi_5^*)(\Phi_4 - \Phi_6)\Big]\,.
\ee
From Eqs.\ \req{eq:amplitudes} and \req{eq:transform} follows
\be
A_{\perp N}\frac{d\sigma}{dt} \sim 2\kappa_T R_V^g R_V \frac{s-u}{\sqrt {-us}} {\rm Im}
                        \big({\cal H}_{++,++}^g +{\cal H}_{+-,+-}^g\big)\,.
\ee
The NLO subprocess amplitudes, derived in \ci{HKM}, provide
\be
{\rm Im}\Big[{\cal H}^g_{++,++}+{\cal H}^g_{-+,-+}\Big]\=-\als C_F\sqrt{\frac{\sh}{-\uh}}\,
         \Big[-\frac{2\th-\uh}{2\sh} + (1+\frac{\th^2}{-\uh\sh})\ln{\frac{-\th}{\sh}}\Big]\,.
\ee
A non-zero result on $A_{\perp N}$ requires proton helicity flip which is provided by the handbag
approach through $R_T$, and phase differences which are obtained from the NLO corrections.
Exactly the same result is found for the transverse target polarization, $P$, \ci{HKM}.
Predictions for $A_{\perp N}$ are of the order of $10\%$ with rather large uncertainties
because of badly known gluon form factor $R_V^g$. 

An example of an observables for which only corrections of order $\als^2$ lead to a
non-trivial result, is the helicity transfer from the incoming to the outgoing photon
\ci{HKM}
\be
D_{LL} \= 1 +  {\cal O}(\als^2)\,.
\ee
A deviation from 1 requires photon helicity flip which is of order $\als$ in the handbag
approach,  see \req{eq:NLO-flip}. The deviation from 1 due to the NLO photon helicity flip
are tiny, of the order of $1\%$.

One may also consider spin correlations between the final state photon and proton ($C_{ij}$)
or between the final state photon and the initial state proton. These observables are similar
to the $A_{ij}$ and $K_{ij}$ ones. For instance,
\be
C_{LL}\=A_{LL}\,, \qquad C_{LS}\=A_{LS}\,.
\ee

\begin{figure}
\centering
\includegraphics[width=0.45\tw]{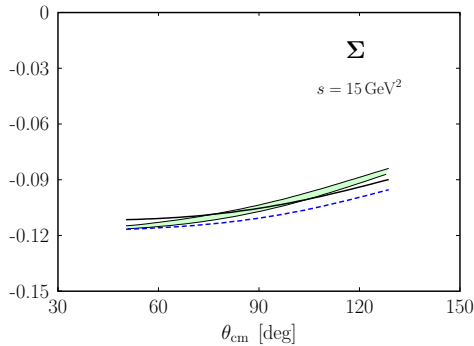}
\caption{\label{fig:Sigma} Left: The photon asymmetry $\Sigma$ versus the c.m.s. scattering
angle at $s=15\,\gev^2$. The shaded band represents the uncertainties due to target mass corrections
for fit \#1 (upper edge: scenario 1, lower edge: scenario 3). Solid (dashed) line: scenario 1, 
fit \#2 (\ci{DK13}).}
\end{figure}

The last observable we want to comment on is the incoming photon asymmetry, $\Sigma$, defined 
as
\be
\Sigma\=\frac{d\sigma(\perp) - d\sigma(\parallel)}{d\sigma(\perp) + d\sigma(\parallel)}
\label{eq:sigma}
\ee
which at least at low energies and/or small $-t$ has been measured or in photoproduction 
of pions, e.g.\ \ci{quinn79}. This observable can be expressed by 
\be
 \Sigma\frac{d\sigma}{dt} \= \frac1{16\pi(s-m^2)^2}\,{\rm Re}\,\Big[
          \big(\Phi_1 + \Phi_5\big)\Phi^*_3 + \big(\Phi_2 - \Phi_6\big)\Phi^*_4\Big]\,.
\ee
It is of order $\als$ and can be expressed by
\ba
\Sigma&=& -\frac{\als}{\pi} C_F\,
         \Big[1 + \frac{t^2}{(s-u)^2}\,\frac{R_A^2}{R_V^2}\frac1{1+\kappa_T^2}\Big]^{-1}\nn\\
      &\simeq& -\frac{\als}{\pi} C_F\,.
\ea
Predictions for $\Sigma$ are depicted in Fig.\ \ref{fig:Sigma} at $s=15\,\gev^2$.  
This asymmetry is only mildly dependent on the Compton form factors.

Before closing this section a remark is in order concerning other models for WACS.
Above we have already mentioned the handbag model proposed by Miller \ci{miller04}. 
In this model massive, point-like quarks are used and the form factors are evaluated
from wave-function overlaps. Due to the quark masses $A_{LL}\neq K_{LL}$ and $D_{LL}$
deviates from unity in the backward hemisphere. In the model invented by the authors 
of \ci{kivel} the leading contribution is the same as our LO result with, however, $R_V=R_A$ and
$R_T=0$. The form factor $R_V$ is not related to GPDs but determined from a fit to
the differential cross section. Corrections to the leading contribution are calculated
from the soft collinear effective theory. This model leads to $A_{LL}\simeq K_{LL}$ with
values similar to our results. Dagaonkar \ci{dagaonkar} proposes an endpoint model
for WACS which bears similarity to the approach discussed in \ci{DFJK1}. In \ci{dagaonkar}
spin effects are not discussed.
 
\section{Summary}
\label{sec:summary}
In the handbag approach the amplitudes for WACS are composed of products of subprocess
amplitudes for Compton scattering off massless quarks and form factors that represent 
$1/x$-moments of GPDs. The relevant GPDs, $H$ and $E$, for the form factors $R_V$ and
$R_T$, are reasonably well known for valence quarks from an analysis of the nucleon
form factors \ci{DK13}. In consequence of the scarce experimental information available
for the isovector axial form factor of the nucleon the GPD $\widetilde H$ and, hence,
the form factor $R_A$, is poorly known. Therefore, predictions on the spin correlations
which are sensitive to $R_A$, on which the interest is focused in this work,
suffer from large uncertainties, in particular for $s$ smaller than about $10\,\gev^2$
where also proton mass corrections of kinematical and dynamical origin are rather large.

Therefore, one may turn around the strategy and extract $R_A$ from WACS data at
sufficiently large Mandelstam variables, as for 
instance from the spin correlations $A_{LL}$ or $K_{LL}$, and use the results as an 
additional constraint in the analysis of $\widetilde H$ with the help of the sum rule for 
the axial form factor. This constraint will also improve the flavor separation of $\widetilde H$.
With only data on $F_A$ on disposal the flavor separation requires assumptions.
Measurements of spin correlations in photoproduction of pions may provide further
constraints on $\widetilde H$. Results similar to \req{eq:ALL}, \req{eq:KLS} hold
for photoproduction \ci{HK00,passek} with, of course, flavor compositions of the form
factors that differ from \req{eq:compton-FF}. Such an analysis will likely improve our
knowledge of $\widetilde H$ for valence quarks at large $-t$ substantially and allow for
a reliable investigation of the impact-parameter distribution of valence quarks with definite 
helicity. \\  
 
{\it Acknowledgements:} The author would like to thank Bogdan Wojtsekhowski 
and Dustin Keller for discussions on the spin correlation observables.

\begin{appendix}
\section{The c.m.s. convention}
\label{sec:HKM}
Here, in this appendix we define various polarization states of the involved particles 
in the c.m. system. We define a unit vector perpendicular to the scattering plane by
\be
{\bf N}\=\frac{{\bf p}\times {\bf p}'}{|{\bf p}\times {\bf p}'|} \=
         \frac{{\bf q}\times {\bf q}'}{|{\bf q}\times {\bf q}'|} 
\label{eq:N}
\ee
where the momenta denote the 3-momenta of the particles involved. As longitudinal, $L$, and
sideway, $S$, spin directions we define 
\be
{\bf L}^{(')}\=\frac{{\bf p}^{(')}}{|{\bf p}^{(')}|}\,, \qquad
{\bf S}^{(')}\=\frac{{\bf N}\times {\bf L}^{(')}}{|{\bf N}\times {\bf L}^{(')}|}\,.
\label{eq:L-S}
\ee
for the incoming and outgoing protons. The vectors ${\bf N}$, ${\bf L}^{(')}$ and  
${\bf S}^{(')}$ form a right-handed system. The $L$ and $S$ directions for the photons 
are defined analogously.   

We use the convention advocated for by Bourrely, Leader and Soffer \ci{BLS} and
define the rotation of a vector through an azimuthal angle $\varphi$ and a polar angle 
$\vartheta$ by the matrix $R(\varphi,\vartheta,0)$ with the Pauli matrices, $\sigma_i$ for 
the protons and the spin-1 matrices for photons and momenta. The different polarization states
of the proton - $L$, $S$ and $N$ - are defined as spin eigenstates of 
${\bf A}\cdot {\bf \sigma}$ where ${\bf A}$ is one of the unit vectors \req{eq:N} and 
\req{eq:L-S}. For the $L$-type polarizations the eigenstates are just the usual helicity 
states whereas for the $S$-type polarization with the eigenvalue $1/2$ of the operator 
${\bf S}^{(')}\cdot {\bf \sigma}/2$ is
\be
|\rightarrow \rangle \= \frac1{\sqrt{2}} \big[\, |+\rangle - |-\rangle\, \big]\,.
\ee
In terms of helicity amplitudes an amplitude for sideway polarization of the initial state proton
reads
\be
\Phi_{\mu'\nu',\mu \rightarrow} \= \frac1{\sqrt{2}} 
               \Big(\Phi_{\mu'\nu',\mu +} - \Phi_{\mu'\nu',\mu -}\Big)\,.
\label{eq:sideway}
\ee
For the $N$-type polarization with positive and negative eigenvalue of 
${\bf N}\cdot{\bf \sigma}/2$ one finds
\be
\Phi_{\mu'\nu',\mu \uparrow(\downarrow)} \= \frac1{\sqrt{2}} 
               \Big(\Phi_{\mu'\nu',\mu +}\, {}^{\phantom{(}+\phantom{)}}_{(-)} \,i\, 
                                    \Phi_{\mu'\nu',\mu -}\Big)\,.
\label{eq:normal}
\ee
Analogously relations hold for the final state proton.

For the photons the $N$ and $S$-type polarization correspond to linear photon polarizations.
They are usually denoted by $\perp$ and $\parallel$, respectively. For the initial state
photon the amplitudes for linear photon polarization read
\ba
\Phi_{\mu'\nu',\perp \nu} &=& \frac{i}{\sqrt{2}} 
               \Big(\Phi_{\mu'\nu',+\nu} + \Phi_{\mu'\nu',-\nu}\Big)\,, \nn\\
\Phi_{\mu'\nu',\parallel \nu} &=& \frac{-1}{\sqrt{2}} 
               \Big(\Phi_{\mu'\nu',+\nu} - \Phi_{\mu'\nu',-\nu}\Big)\,.
\label{eq:linear}
\ea
Again, analogous relations hold for the final state photon.

\section{The lab system convention}
\label{sec:BGLMN}
In \ci{BGLMN} the lab system is considered for the definition of spin directions.
For both the photons as well as for the final state proton spin directions are 
defined which, after boosting from the lab system to the c.m. system, fall together 
with our conventions, see \req{eq:N}, \req{eq:L-S}. However, for the initial state proton, 
being at rest in the Lab system, the same spin directions as for the initial state photon 
are chosen in \ci{BGLMN}. After a boost to the c.m. system one sees that this choice 
implies differences as compared to \ci{HKM}:
\be
{\bf L}^{BGLMN}\=-{\bf L}^{HKM}\,, \quad {\bf S}^{BGLMN}\=-{\bf S}^{HKM}\,,
\quad {\bf N}^{BGLMN}\={\bf N}^{HKM}\,.
\ee
Suppose the scattering plane is the ${\bf e}_1 - {\bf e}_3$ plane and ${\bf N}={\bf e}_2$.
Then the corresponding spin states for the ${\bf L}$ and ${\bf S}$ directions are the 
eigenstates of $\sigma_3/2$ and $\sigma_1/2$ respectively instead of $-\sigma_3/2$ and 
$-\sigma_1/2$ as is the case for the conventions discussed in App.\ \ref{sec:HKM}. In terms 
of proton helicity the spin state with the eigenvalue $1/2$ of the operator $\sigma_3/2$ 
corresponds to negative helicity. For the sideway polarization with eigenvalue $1/2$ of 
the operator $\sigma_1/2$ is
\be
|\rightarrow \rangle^{BGLMN} \= \frac1{\sqrt{2}} \big[\, |+\rangle + |-\rangle\, \big]\,.
\ee
For a helicity amplitude this implies
\be
\Phi^{BGLMN}_{\mu'\nu'\mu \rightarrow} \= \frac1{\sqrt{2}} 
               \Big(\Phi_{\mu'\nu'\mu +} + \Phi_{\mu'\nu'\mu -}\Big)
\ee
 instead of \req{eq:sideway}. Hence,
\be
A_{LL}^{BGLMN}\=-A_{LL}^{HKM}\,, \qquad A_{LS}^{BGLMN}\=-A_{LS}^{HKM}
\ee
but
\be
K_{LL}^{BGLMN}\=\phantom{-}K_{LL}^{HKM}\,, \qquad K_{LS}^{BGLMN}\=\phantom{-}K_{LS}^{HKM}
\ee

\end{appendix}

\end{document}